\documentclass[11pt, a4paper]{article}
\pdfoutput=1

\usepackage{graphicx}
\usepackage{psfrag}
\usepackage{enumerate}
\usepackage{amssymb}
\usepackage{jheppubcustom}

\newcommand{\be}{\begin{equation}}
\newcommand{\ee}{\end{equation}}
\newcommand{\beq}{\begin{eqnarray}}
\newcommand{\eeq}{\end{eqnarray}}

\def\[{\left [}
\def\]{\right ]}
\def\({\left (}
\def\){\right )}

\def\r2{\sqrt{2}}

\def\xsi{\xi}

 \def\simleq{\; \raise0.3ex\hbox{$<$\kern-0.75em
      \raise-1.1ex\hbox{$\sim$}}\; }
   \def\simgeq{\; \raise0.3ex\hbox{$>$\kern-0.75em
      \raise-1.1ex\hbox{$\sim$}}\; }

\newcommand{\bbibitem}[1]{\bibitem{#1}\marginpar{#1}}

\newcommand{\figref}[1]{Fig. \ref{#1}}
\newcommand{\secref}[1]{Sec. \ref{#1}}

\def\Label#1{\label{#1}%
  \smash{\hbox to0pt{\raise1ex\hbox{\tiny[#1]}\hss}}}
\def\noLabels{\let\Label=\label}
\def\nobbibitem{\let\bbibitem=\bibitem}

\newcommand{\dph}{\delta\phi}

\begin{document}

\noLabels
\nobbibitem

\DeclareGraphicsExtensions{.pdf,.png,.gif,.jpg,.eps}

\title{Analyzing Cosmic Bubble Collisions}

\author{Roberto Gobbetti and Matthew Kleban}

\emailAdd{rg1509@nyu.edu}
\emailAdd{mk161@nyu.edu}

\affiliation{\it Center for Cosmology and Particle Physics\\
New York University \\
New York, NY 10003, USA}


\abstract{We develop a set of controlled, analytic approximations to study the effects of bubble collisions on cosmology.  We expand the initial perturbation to the inflaton field caused by the collision in a general power series, and determine its time evolution during inflation in terms of the coefficients in the expansion.  In models where the observer's bubble undergoes sufficient slow-roll inflation to solve the flatness problem, in the thin wall limit only one coefficient in the expansion  is relevant to observational cosmology, allowing nearly model-independent predictions.  We discuss two approaches to determining the initial perturbation to the inflaton and the implications for the sign of the effect (a hot or cold spot on the Cosmic Microwave Background temperature map).  Lastly, we analyze the effects of collisions with thick-wall bubbles, \emph{i.e.} away from the thin-wall limit.}

\maketitle

\section{Introduction}

The physics of cosmic bubble collisions---collisions between bubbles formed from first-order phase transitions in models that undergo false-vacuum eternal inflation---has recently attracted significant attention \cite{Freivogel:2005vv, ggv, ben, Aguirre:2007an, wwc,Aguirre:2007wm, wwc2, bubmeas, Aguirre:2008wy, ktflow,Easther:2009ft, Giblin:2010bd, Czech:2010rg, Salem:2010mi, Johnson:2011wt}.  String theory predicts the existence of a large number of meta-stable de Sitter vacua  \cite{discretum, lennylandscape, kklt}.  Most of these vacua inflate extremely rapidly, at a rate set by the string scale.  In a few, the vacuum energy is expected to be comparable to the measured dark energy.  Hence, in this model the cosmological constant problem is solved in the sense that the vacuum energy is an environmental variable that varies from place to place in the universe, and in some places it is consistent with the measured value.

Our observable universe has a very small vacuum energy.  In a model that undergoes false-vacuum eternal inflation, the most natural origin for such a region is from a bubble that nucleated in a first-order phase transition from a (presumably rapidly inflating) parent false vacuum.  If so, the observable universe fits inside a bubble surrounded by a parent false vacuum---a parent vacuum that is evidently unstable to the nucleation of bubbles.  Due to this instability, other bubbles are guaranteed to eventually form in the false vacuum surrounding ours.  Some of these bubbles will collide with ours, making a potentially observable imprint on cosmology.  

Much of the recent activity in this area has focused on quantifying the effects of  collisions on cosmological observables.  Significant progress has been made on that front \cite{wwc2, Czech:2010rg, Kleban:2011yc}.  The results obtained thus far took as a starting point a particular form for the perturbation at the end of inflation that was first derived in \cite{wwc2} using various approximations, and
 the WMAP data set was recently analyzed using the predictions of \cite{wwc2} as a template \cite{Feeney:2010jj}.  
 
The main result of this paper is a model independent prediction for the collision perturbation in the early universe (\secref{powerseries}).   Using the symmetries of the bubble collision as a tool, we demonstrate analytically that for a generic perturbation at the beginning of inflation that is consistent with the symmetries, the main effect by the end of inflation is indeed the one considered in previous work---a curvature perturbation to the early universe at reheating that is piecewise linear, but that has a sharp change in its first derivative (a ``kink'') in the direction towards the collision bubble at the edge of the region affected by the collision.  The kink is sharp in the thin wall limit, and we quantify how it smooths out if the collision involves bubbles with thick walls.   

The enormous expansion during the period of slow-roll inflation in the observer's bubble after the collision substantially affects the form of the perturbation and gives rise to this prediction.  By the end of inflation, a ``step" discontinuity present at the start of inflation evolves into a kink (a discontinuity in the derivative).  It is therefore the sign and magnitude of the initial step discontinuity that is the single most important factor in determining the cosmological signatures.  An analytical approach to this is essential, because---at least for the moment---the numerical simulations of bubble collisions that have been done \cite{Easther:2009ft, Giblin:2010bd, Johnson:2011wt} cannot incorporate the full period of slow roll inflation.

The paper is organized as follows.  In  \secref{background} we very briefly review the physics of bubble collisions, and discuss the approximations and assumptions used in analyzing their effects.  In \secref{powerseries} we develop a general technique for analyzing the effects of a bubble collision before and during inflation.  in  \secref{examples} we apply our techniques to several specific examples.  In  \secref{thickwall} we analyze the effects of collisions involving bubbles with walls of finite thickness, and in  \secref{conclusions} we conclude.

\section{Background} \label{background}

Before taking collisions into account, bubbles that form during a first-order phase transition from a false vacuum state will be highly symmetric.  The reason for this is twofold:  the near de Sitter invariance of the false vacuum state, which (since the false vacuum inflates) should be very close to perfect, and the fact that the dominant instanton in such a de Sitter background is expected to be invariant under rotations in the four Euclidean signature dimensions, that is under an $SO(4)$ group of isometries.  As a result of this symmetry, each unperturbed bubble contains a spatially infinite, homogeneous and isotropic Freidmann-Robertson-Walker (FRW) cosmology with negative spatial curvature (with an $SO(3,1)$ isometry group).  Note that none of these conclusions rely on the thin-wall approximation.

For this cosmology to be consistent with observational data, a period of slow-roll inflation is required to solve the flatness problem and produce a scale-invariant spectrum of perturbations \cite{Freivogel:2005vv}.  The ``big bang" of the bubble---FRW time $t=0$, where the scale factor $a(0)=0$---is a null surface that coincides with the future lightcone of the center of the bubble at the moment it appears.  If the physics of the phase inside the bubble is appropriate, slow-roll inflation takes place along negatively curved time-slices at times $t>0$; therefore, cosmic bubble collisions occur \emph{before} slow-roll inflation (\figref{delta}).

A collision between two such bubbles generically makes a large perturbation to the ``big bang" surface $t=0$.  Hence, the problem of determining the effects of a single collision on cosmology is the problem of evolving an inhomogeneous and anisotropic big bang forward in time, which is generally very difficult.  Fortunately, the high degree of symmetry preserved by the collision is of help.  By a de Sitter boost, one can always choose a frame in which the two bubbles nucleate simultaneously.  In that frame, one expects rotation invariance about the axis that connects the centers of the two bubbles, as well as boost invariance in the two directions transverse to that axis.  In other words, one expects the spacetime and field configuration of a 2-bubble collision to preserve an $SO(2,1)$ group of isometries  (see \emph{e.g.} \cite{Hawking:1982ga, ben, wwc, Kleban:2011pg}).

In addition, in any model consistent with current observations, the effects of the collision must be weak and can at some point in the evolution be treated perturbatively.  Slow-roll inflation must have taken place everywhere in the past lightcone of the earth today.  As we will see, this significantly simplifies the problem.

As we will explain, one can choose coordinates in which the collision between the two bubbles takes place at one instant everywhere along a hyperboloid.  In the thin wall approximation discussed below, the effects of the collision are confined to the causal future of this hyperbolic surface at that instant.  We refer to the (null) boundary of this region as the ``collision lightsheet."

\subsection{Approximations} \label{approxsec}

In order to analyze the effects of bubble collisions on observational cosmology, past analyses used several approximations and assumptions that we detail here.  

\paragraph{Effective field theory:}  The collision can be analyzed using a relatively simple effective field theory of scalar fields in $3+1$ spacetime dimensions.  This approximation can be violated for example in models where the tunneling is from lower dimensions into 3+1 \cite{Graham:2010hh, BlancoPillado:2010uw, Adamek:2010sg}, and its validity in realistic string compactifications has yet to be firmly established.

\paragraph{Linearity:}  As we will see, the effects of a single collision (\emph{i.e.} a collision between two bubbles) are fairly easy to determine.  Multiple collisions can be analyzed as well provided that the interaction between them is weak; that is, provided that by the time the collision lightsheets overlap the effects are in the perturbative regime.  

\paragraph{Dilute bubbles:} Prior to any collisions, each bubble is invariant under an $SO(3,1)$ isometry group.  The $SO(3,1)$ invariance holds for bubbles described by the Coleman-de Luccia instanton even away from the thin wall limit, as well as for Brown-Teitelboim membrane formation \cite{Brown:1988kg} and the Hawking-Moss instanton \cite{Hawking:1981fz}.  

\paragraph{Symmetry breaking:} The collision breaks the $SO(3,1)$ symmetry to an $SO(2,1)$ subgroup (rotation and ``boosts" in the plane transverse to the axis connecting the centers of the two bubbles).  This is the case in simple models for collisions between Coleman-de Luccia bubbles, and is supported by numerical simulations \cite{Easther:2009ft}.

\paragraph{Coupling to the inflaton:} The collision perturbs the inflaton field, which is a scalar or effective scalar mode that drives slow-roll inflation in the bubble after it forms.  While our analysis does not depend on the specifics of the coupling in any way, it focuses on single-field inflation and ignores the effects of the collision on other fields that might be of cosmological interest.

\paragraph{Thin wall:} The collision bubble is ``thin wall", meaning that the effects of the collision are confined within  a sharply defined spacetime region.  We will investigate the consequences of relaxing this assumption in Section \ref{thickwall}.

\paragraph{Small curvature:}  In the most convenient set of coordinates, each bubble contains a negatively curved FRW cosmology.  A single collision produces a ``cosmic wake" \cite{Kleban:2011yc} that sweeps across the volume of the bubble.  The wake at any given time is bounded by a hyperboloid, which previous analyses approximated as planar \cite{wwc2,Kleban:2011yc}. As we demonstrate in Appendix \ref{curvature}, this approximation is valid for our universe, or in general when there has been enough slow-roll inflation to solve the flatness problem.

\section{The collision perturbation}\label{powerseries}

 A single bubble, before taking collisions into account, contains a homogeneous and isotropic negatively curved FRW cosmology.  Before inflation, that is to say at FRW times $0<\tau<1/H_i$ (where $H_i$ is the Hubble parameter during slow-roll inflation) the spacetime inside a bubble  is dominated by the negative spatial curvature, and can therefore be well-approximated as flat Minkowski space sliced with hyperbolic 3-space (eq. \eqref{H3} in the limit $\tau H_i \ll 1$).  For times $\tau>1/H_i$ the effects of the inflaton potential become dominant, and for the duration of slow-roll inflation the spacetime is de Sitter up to standard slow-roll corrections.  

Immediately after a collision, the effects on both the spacetime geometry and the inflaton field may be large.  However, if slow-roll inflation begins anywhere in the region affected by the collision, the effects of the collision rapidly inflate away and become perturbative. Given the lack of large fluctuations in the Cosmic Microwave Background (CMB), we know that all the space inside our past lightcone underwent slow-roll inflation.  Therefore the effects of any collision that is potentially visible today must have become small enough at some time to allow inflation to begin everywhere within our past lightcone, and from that point in time onward they can be treated perturbatively.  This is the key to our analysis.

It is important to note that if the inflationary epoch lasted significantly longer than the observational bound on curvature requires (number of e-folds $N\gtrsim62$, with the number depending logarithmically on various factors such as the reheating temperature), a bubble collision would inflate away into a negligible perturbation in the CMB  and would not be detected. On the other hand, at least in some simple models \cite{Freivogel:2005vv}, the number of e-folds of inflation is not expected to greatly exceed the lower bound set by observation.

\begin{figure}
	\begin{center}
	\includegraphics[angle=90, width=1.1 \textwidth]{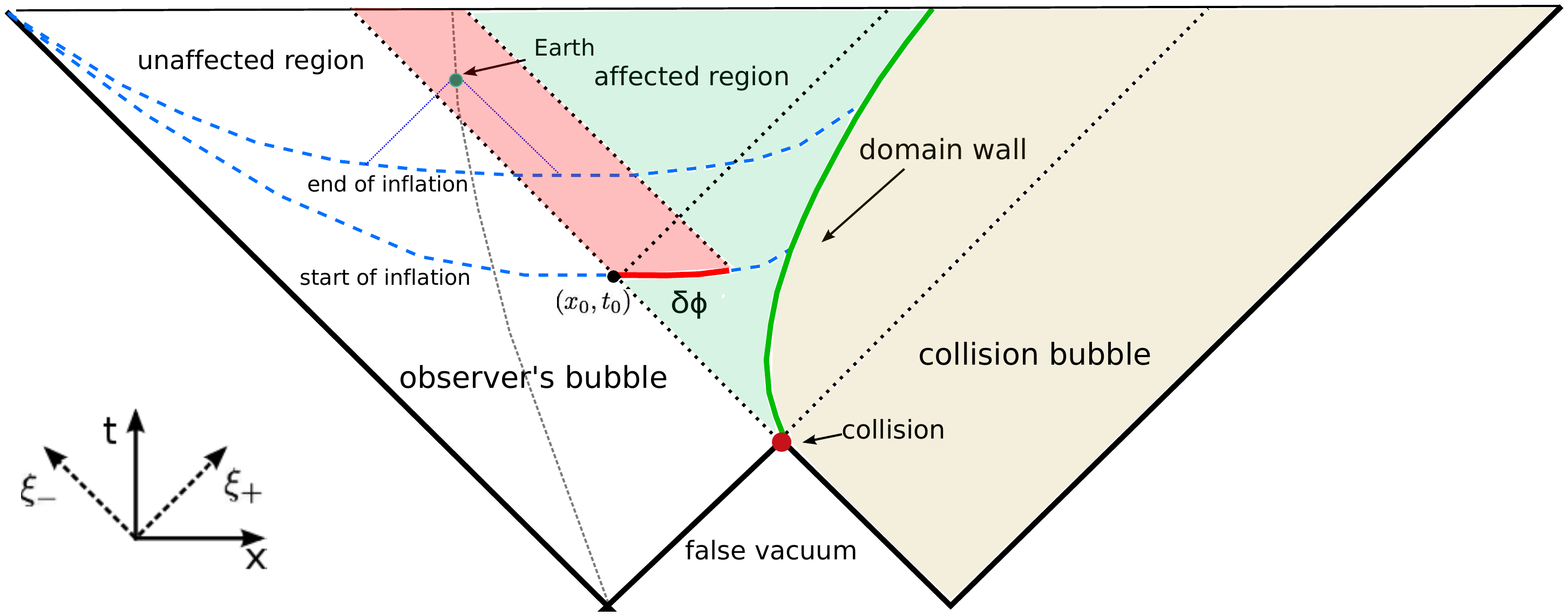}
	\caption{Spacetime diagram of a cosmic bubble collision.  Each point in the plane of the diagram is a two dimensional hyperbolic surface, with a radius of curvature that varies with position in the plane.   The left-most dashed line bounds the region in the observer's bubble that is affected by the collision; referred to in the text as the collision lightsheet.  Knowledge of the perturbation $\delta\phi$ and its derivative provides initial conditions for cosmological evolution in the (red) shaded region.}
	\label{delta}
	\end{center}
\end{figure}

We will refer to the time when the effects of the collision become perturbative everywhere in our past lightcone as $t_{0}$.   Knowledge of the inflaton perturbation $\delta \phi$ and its first time derivative at $t_0$ in this volume provides initial conditions that  determine the later time evolution of all cosmological observables (see Fig. \ref{delta}).

According to our assumption regarding symmetry breaking, the collision perturbation preserves an $SO(2,1)$ isometry group.  During slow-roll inflation, the spacetime is approximately de Sitter.  A convenient slicing of de Sitter spacetime that preserves a manifest $SO(2,1)$ isometry is:
\beq \label{hypmet}
ds^2 = {-dt^2 \over h(t)}+h(t)\, dx^2 + t^2 dH_2^2 ,%
\eeq%
where%
\beq%
h(t) = 1+ (H_i t)^{2}, \label{hdef}
\eeq%
where $H_i$ is the inflationary Hubble scale and
$dH_2 ^2 = d\rho^2 + \sinh^2 \rho d\varphi^2$ is the metric on a hyperboloid of unit curvature.

\subsection{Taylor expansion near the lightsheet} \label{Taylor}

Without a model for the microphysics of the collision, we cannot determine the initial perturbation $\delta \phi, \dot {\delta \phi}$ at $t=t_{0}$.  Instead, we will simply expand the initial perturbation in a general power series and then solve for its time evolution.  As we will see, the late-time effects are dominated by a single term in the expansion of the initial conditions---the term representing a discontinuity in $\delta \phi$ across the lightsheet.

To lowest order in the slow-roll expansion, perturbations $\delta \phi$ to the inflaton field satisfy a free, massless wave equation in de Sitter space \cite{mfb}:
\beq%
\Box \phi = -{1\over t^2} \partial_t \left[ t^2 (1+(H_i t)^2) \partial_t \phi\right] +{1\over 1+(H_i t)^2} \partial^2_x \phi = 0 \label{diffeqph}.%
\eeq%
  It turns out that this equation can be solved in full generality assuming $H_2$ symmetry, and the solution is (\cite{wwc}):
\beq \label{hypsol}
\delta \phi(t,x) &=& f(\xi_{-})-{1\over t} f'(\xi_{-})+g(\xi_{+}) +{1\over  t} g'(\xi_{+}),%
\eeq%
where $f$ and $g$ are arbitrary functions of one variable,  primes denote derivatives with respect to the argument, $\xi_{\pm} =x\pm \eta$ are null coordinates, and 
\be
\eta = \int dt/h(t)= H_i^{-1} \tan^{-1}(H_i t) - \pi/(2 H_i)
\ee
 is the conformal time.  Inflation proceeds for $N\sim60$ efolds \cite{ll}, after which the coordinate $t\sim H_i^{-1} e^{N}$ is very large and $\eta \sim -(H_{i}^{2} t)^{-1} \sim 0$ is exponentially small.  After a few efolds, to a good approximation one can drop the ``1''s in \eqref{hdef}, and \eqref{hypsol} becomes
\beq \label{pert}
\delta \phi(t,x) \approx f(\xi_{-}) + H_i^2 \eta f'(\xi_{-})+g(\xi_{+})-H_i^2 \eta g'(\xi_{+}),%
\eeq%
where again $\xi_{\pm} \approx x \pm \eta$, $\eta \approx  -1/(H_i^2 t)$ is the conformal time.  This approximation is equivalent to neglecting the effects of spatial curvature on the perturbation and is justified in detail in Appendix \ref{curvature}.

To proceed, we wish to describe the perturbation $\delta \phi$ in a region near the edge of the collision lightsheet, at the time $t_0$ near the beginning of inflation.
As mentioned above, in the thin-wall approximation the effects of the collision are entirely confined behind a lightsheet: at the initial time $t=t_{0}$, the perturbation $\delta \phi, {\delta \dot  \phi} \sim \theta(x-x_{0})$, where $x= x_{0}$ is the location of the lightsheet at $t=t_{0}$ (\figref{delta}).  (We will explore the consequences of relaxing the thin-wall approximation in Section \ref{thickwall}.)

Rather than expanding $\delta \phi,{\delta \dot  \phi}$ at $t=t_{0}$, it is more convenient to expand the functions $f, g$ of \eqref{hypsol}.  These functions contain the same information as $\delta \phi$ and $\delta \dot \phi$.  In the thin wall approximation they are proportional to a step function that is non-zero only inside the collision lightsheet:
\beq
	f(x-\eta_{0}) &=& M\sum_{n=1}^{\infty}a_nH_{i}^{n}(x- x_{0})^n\theta(x- x_{0}) \label{dphf0} \\
	g(x+\eta_{0}) &=&M \sum_{n=1}^{\infty}b_nH_{i}^{n}(x- x_{0})^n\theta(x- x_{0}) \label{dphg0}.
\eeq
Here $a_{n}$ and $b_{n}$ are dimensionless coefficients, $M$ is a parameter with dimensions of mass, and $\eta_0 = \eta(t_0)$ is the conformal time corresponding to $t=t_0$.  The sum begins with $n=1$, as an $n=0$ term with non-zero coefficient would imply an infinite value for the perturbation $\delta \phi$ \eqref{pert}.  Written in this form, it is trivial to extend to the full time-dependence:
\beq \label{exp}
	f(x-\eta) &=& M\sum_{n=1}^{\infty}a_nH_{i}^{n}(x- x_{0}-\eta+\eta_0)^n\theta(x- x_{0}-\eta+\eta_0) \label{dphf} \\
	g(x+\eta) &=& M\sum_{n=1}^{\infty}b_nH_{i}^{n}(x- x_{0}+\eta-\eta_0)^n\theta(x- x_{0}+\eta-\eta_0) \label{dphg}.  
\eeq

These expansions determine $\delta \phi (\eta, x)$.  After many efolds of inflation when $\eta \sim 0$, $g$ is non-zero everywhere inside the lightsheet, $x- x_{0}>\eta_{0}=-|\eta_{0}|$.  By contrast, the right-moving mode $f$ is non-zero only in the region of spacetime $x- x_{0}>-\eta_{0}=|\eta_{0}|$, a coordinate distance $\Delta x = 2 |\eta_0|$ from the edge of the lightsheet.  Physically, $f$ corresponds to a component of the perturbation $\delta \phi$ that moves directly away from the observer at the speed of light, starting from early in inflation (illustrated in \figref{delta} by the dashed line in the $\xi_+$ direction originating from the point $(x_0, t_0)$).  As such it is unobservable, at least in a situation in which the collision lightsheet intersects the observer's last scattering volume.   Therefore,
in computing the observable effects of a bubble collision, we can neglect the rightmoving modes $f(\xi_-)$.

As can be seen from \eqref{exp}, in the vicinity of the lightsheet at late times ($x \sim  x_{0}+\eta_{0}, \eta \sim 0$) the dominant term in \eqref{exp} is the one proportional to $b_{1}$ (barring  a large hierarchy among the dimensionless coefficients $b_{n}$).  Longer inflation suppresses the higher terms exponentially, since it reduces the range $\Delta x$ within the observer's last scattering volume by a factor that scales with $e^{-N}$ (it also suppresses the effects of the term proportional to $b_1$---but by the lowest power if $e^{-N}$).

Therefore, generically the late-time perturbation is
\be 
\delta \phi (\eta, x) = M H_{i} b_{1} x \theta(x) + {\cal O}(\eta, x^{2}),
\ee
where for notational convenience we have chosen the origin of $x$ so that $x_{0}=-\eta_{0}$ and the collision lightsheet is at $x=0$ when $\eta=0$.

To relate this to the perturbation at early times $\eta = \eta_{0}$, note that $g(\xi_{+}) = b_{1} H_{i} \xi_{+} \theta(\xi_{+})$ corresponds to a perturbation $\delta \phi(x, \eta_{0}) = M H_{i} b_{1} x \theta(x+\eta_{0})$.  This contributes a discontinuity 
\be
\Delta\phi = M H_{i} b_{1} |\eta_{0}|
\ee
 in the inflaton field value at the edge of the collision lightsheet.  Hence, the amplitude of the ``left-moving part'' of the discontinuity in $\delta \phi$ at the edge of the lightsheet is what determines the coefficient of the $x\theta(x)$ kink at late times (this same conclusion is derived in a slightly more direct way in Appendix \ref{seriesappend}).  The parameter $M b_1$, along with the angular size of the affected disk and its position on the CMB sky, fully parametrize the leading effects of a thin-wall bubble collision.

Our conclusion is that a general initial inflaton perturbation evolves to a perturbation that has been smoothed essentially by integrating in $x$ once (see Appendix \ref{seriesappend}).   In other words, an initial discontinuity becomes a kink, a kink becomes a function with discontinuous second derivative, etc.  This is in accord intuitively with the more standard treatment of inflationary perturbations in $k$-space, where $\delta \phi \sim k^{-3/2}(1 \pm i \eta k)$.  Near the end of inflation $|\eta k| \ll 1$,  and so an initial perturbation in Fourier space evolves to one with one less power of $k$.

\section{Examples}\label{examples}

As we have seen, the dominant effect of a bubble collision arises when the collision creates a discontinuity in the inflaton field across the collision lightsheet (when $b_1$ in \eqref{dphg0} is not zero).  When present near the beginning of slow-roll inflation, such a discontinuity in the field ``inflates away'' into a discontinuity in the first derivative of the field ($x \theta(x)$), which was the starting point for various analyses of the effects on cosmological observables \cite{wwc, Larjo:2009mt, Czech:2010rg, Kleban:2011yc}.

In this section we discuss two approaches in which we can analytically study the effects of the collision on the universe before inflation.  While either approach could in certain microphysical models be used to determine the coefficients $a_n$ and $b_n$ in the initial perturbation to the inflaton, doing so is not our primary purpose.  Instead, we wish to show that the  effect of a bubble collision on the inflaton across a broad class of models is in fact to create a discontinuity in its value across the lighsheet (in the thin wall approximation)---in other words, we wish to show that $b_1 \neq 0$ generically, rather than calculate its value in any specific model.

Our first approach makes use of the ``free passage'' approximation of \cite{Giblin:2010bd}, while the second treats the domain wall between the bubbles as a boundary condition on the inflaton.  In a sense, these two approaches constitute opposite extremes---free passage applies when the kinetic energies are so large that the fields can be treated as free, and the domain wall between the bubbles is irrelevant or even non-existent, while treating the domain wall as a boundary condition requires strong non-linearities in the field.  Nevertheless, in both cases, the collision generically produces a discontinuity in the inflaton across the collision lightsheet.  In addition, numerical simulations support the conclusion that thin-wall bubble collisions lead to discontinuities in the field value along the lightsheet of the collision \cite{Aguirre:2008wy, Easther:2009ft, Giblin:2010bd, Johnson:2011wt}.

\subsection{Free passage}

\begin{figure}
\begin{center}$
\begin{array}{| c| c |}
\hline
	\includegraphics[angle=90, width=0.5\textwidth]{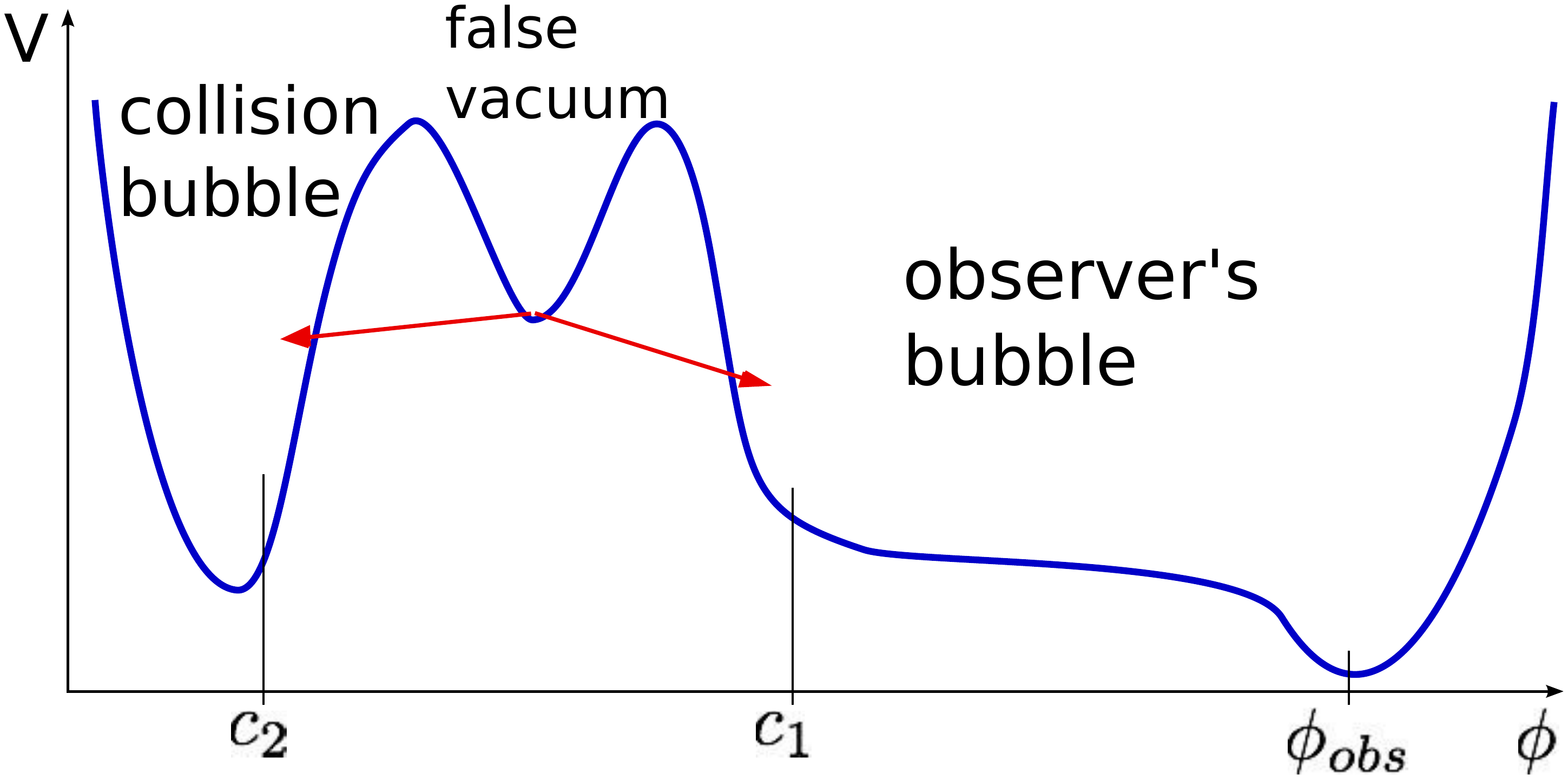} & \includegraphics[angle=90, width=0.5\textwidth]{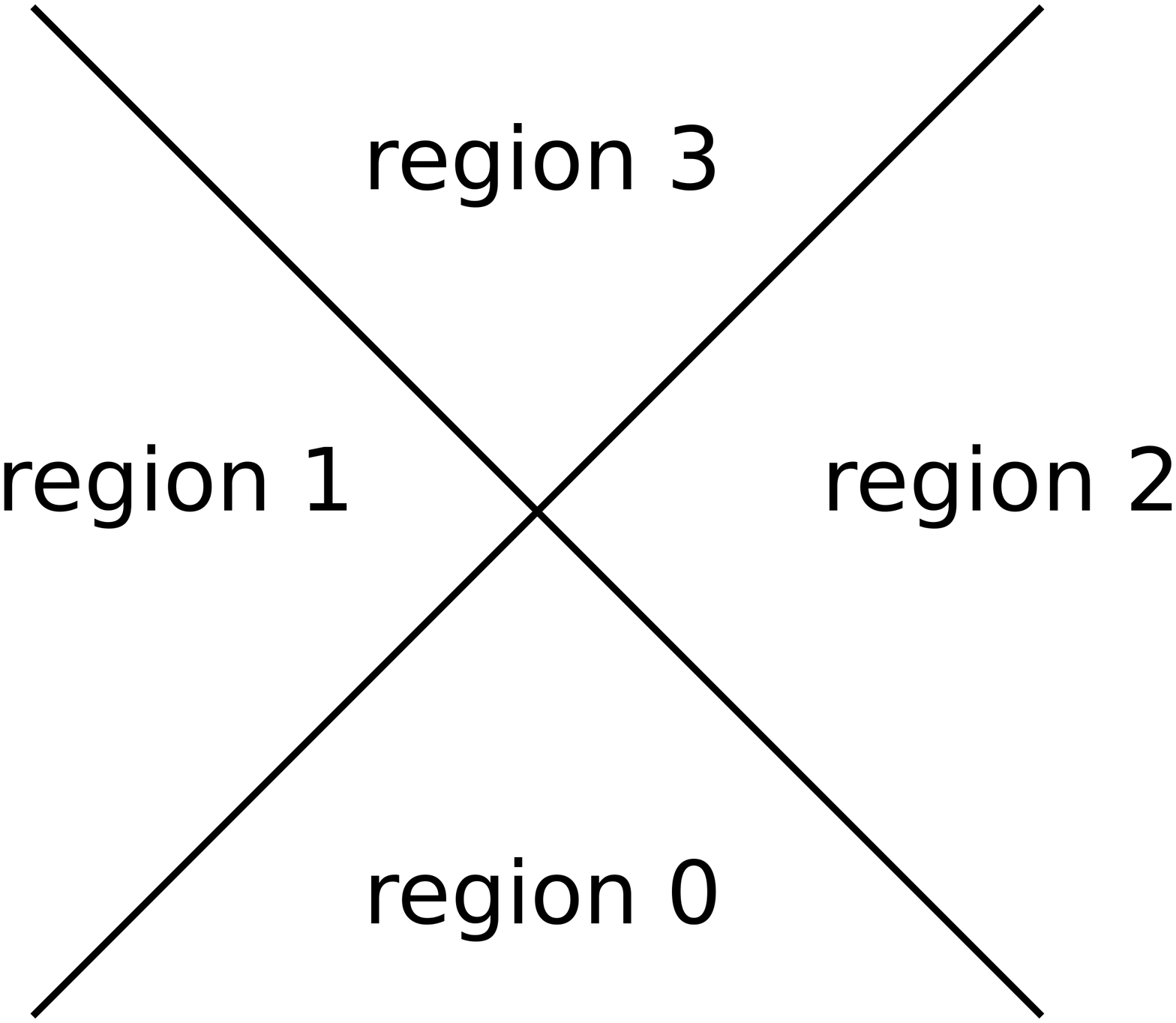} \\
		\hline
\end{array}$
\end{center}
\caption{\label{landscape} Left panel: A potential $V(\phi)$ with three minima.  The Coleman-de Luccia tunneling paths from the central false vacuum are indicated with arrows.  Bubbles that tunnel to the right will inflate along the slow-roll plateau leading down into the observation bubble's minimum. Right panel: Schematic of the field configuration prior to the collision.}
\end{figure}

When two planar scalar-field domain walls collide head-on, if the kinetic energy of the walls exceeds the relevant energy differences in the scalar potential, the  field configuration immediately after the collision is determined in a very simple way by the pre-collision field values  \cite{Giblin:2010bd} (see Appendix A of \cite{Kleban:2011cs} for an extension to collapsing spherical configurations that ``turn inside out").  
Consider a model with several scalar fields $\vec \phi$ and a potential $V(\vec \phi)$.  Initially,  a region of $\vec \phi=\vec \phi_{0}$ separates two regions where  $\vec \phi = \vec \phi_{1}$ and $\vec\phi = \vec \phi_{2}$ bounded by ingoing planar walls (see \figref{landscape}).  The results of  \cite{Giblin:2010bd} show that immediately after the walls collide, the field value in region 3  is
\be \label{freepass}
	\vec \phi_3=\vec \phi_1+\vec \phi_2-\vec \phi_{0}.
\ee
In the case of free fields ($V(\vec \phi)$ constant) in flat Minkowski spacetime, the field in region 3 will remain fixed and equal to $\vec \phi_{3}$ for all times after the collision.  The same follows in interacting models if $\vec \phi_{3}$ happens to be a minimum of the potential.  In general, the post-collision evolution of the field in region 3 depends on the model---given  $\vec \phi=\vec \phi_{3}$ as an initial condition, the time evolution of $\vec \phi(t)$ will depend
on the scalar potential $V(\vec \phi)$, as well as the spacetime curvature.

We wish to consider a collision between two thin-wall bubbles containing field values $\vec \phi_{1}$ and $\vec \phi_{2}$, embedded in a rapidly inflating parent vacuum $\vec \phi_{0}$.  As before we will ignore the hyperbolic curvature of the walls (although the analysis of \cite{Kleban:2011cs} can be extended to that case).  Bubble 1 is the observation bubble, and we will assume $\vec \phi_{1}$ falls on or in the basin of attraction of a slow-roll inflationary slope that ends with an observationally acceptable vacuum.  Bubble 2 is the collision bubble, and $\vec \phi_{2}$ may be a minimum or close to a minimum of the potential.

We would like to apply this approach to a model that is potentially consistent with observation.  This requires the observer's bubble to undergo a period of slow-roll inflation.  A simple such model is the single-field model illustrated in \figref{landscape}, where the observer's minimum $\phi=\phi_{obs}$ is separated from the parent vacuum by a long slow-roll plateau.  After tunneling,  the fields near the bubble walls are not in the minimum.  Instead, the tunneling point---which is the value of the field on the walls of the bubble---is separated in field space from the false vacuum only by the width of the barrier.  Cosmological evolution inside the bubble is rapid at first, but ``friction" due to spatial curvature prevents ``overshoot" and causes the inflaton to move only a distance roughly equal to the barrier width before coming to near rest and then slowly rolling along the plateau \cite{Freivogel:2005vv}.

If the kinetic energy in the walls is large at the time of collision, the free passage result \eqref{freepass} should apply immediately after the collision.  The field value $\vec \phi = \vec \phi_{3}$ may or may not fall on or (in a multi-field model) in the basin of attraction of the slow-roll inflationary plateau of bubble 1.   If it does not, the collision will have a very large effect on the region 3, which is almost certainly not compatible with observation.  Therefore we will assume that it does fall on the plateau, so that standard slow roll inflation takes place inside the affected region.  

In this case the fields in regions 1 and 3 may behave approximately as free fields for at least a time of order the inflationary time $H_{i}^{-1}$ determined by the height of the slow-roll plateau, at which point slow-roll inflation begins.  The analysis of \secref{Taylor} applies directly to this situation, with the initial field values approximately determined by \eqref{freepass}.  We assume that the backreaction of the energy in field gradients is small, so that the spacetime in regions 1 and 3 is de Sitter space with Hubble constant $H_{i}$.

The simple equation \eqref{freepass} must be applied with care to collisions involving such bubbles, because the field inside the bubble is evolving rather than constant.  However, for early times after the collision and for the first few efolds of inflation it should be valid in the vicinity of the tunneling point.  The most appropriate value for the free-passage approximation is one not far from the tunneling point (labelled $c_1$ and $c_2$ in \figref{landscape}).   Numerical simulations are probably required to determine the accuracy of this approximation in any given model.

With $\phi_{0}=0, \phi_{1}=c_{1},$ and $\phi_{2}=c_{2}$ (see \figref{landscape}), the value of the field in the region affected by the collision (region 3 in \figref{landscape}), is (\ref{freepass}):
\be
	\phi_3 \simeq c_1+c_2
\ee
soon after the collision.  The evolution of the field in region 2 is unobservable, but for illustrative purposes in this section we will treat region 2 as de Sitter space with the same vacuum energy as regions 1 and 3.

To find the time-dependence in region 3, we will make use of the left and right moving formalism developed in the first section. Expanding $\phi$ as in (\ref{hypsol}),  we should require that $f$ and $g$, are non-zero only in the regions 1 and 2 respectively, and that both are non-zero in region 3. Additionally, they should satisfy
\beq
	f(z)-\frac{1}{t_c}f'(z)&=&c_2 \\
	g(z)+\frac{1}{t_c}g'(z)&=&c_1
\eeq
in the regions where are non-zero; $(x_c,t_c)$ are the coordinates of the collision point.

The solutions to these equations are
\beq
	f(z)&=&c_2(1-e^{t_cz}) \\
	g(z)&=&c_1(1-e^{-t_cz})
\eeq
and therefore the solution for $\phi$ with the complete causal structure is
\beq
	\phi(x,t) &=& c_2\left(1-e^{t_c(x-\eta-x_c+\eta_c)}+\frac{t_c}{t}e^{t_c(x-\eta-x_c+\eta_c)} \right) \theta (-x+\eta+x_c-\eta_c) +\nonumber \\
	&& c_1\left(1-e^{-t_c(x+\eta-x_c-\eta_c)}+\frac{t_c}{t}e^{-t_c(x+\eta-x_c-\eta_c)} \right) \theta (-x-\eta+x_c+\eta_c) \nonumber \\
	&& \label{phitot}
\eeq
where $\eta=-1/t$ and $\eta_c=-1/t_c$. 
It is straightforward to check that this solution shows the correct behavior around the collision point $(x_c,t_c)$.

At late times $t \gg 1$ (corresponding to many efolds of inflation), the last term in the parenthesis of each line of (\ref{phitot}) becomes negligible, leaving only the $(1-e^{\ldots})$. The leading term of this expression is  linear in $x$, showing again that the late time behavior near the light cone has a discontinuity in the first derivative, but not  in the field value. 

\begin{figure}
\begin{center}$
\begin{array}{| c| c |}
\hline
	\includegraphics[width=0.45\textwidth]{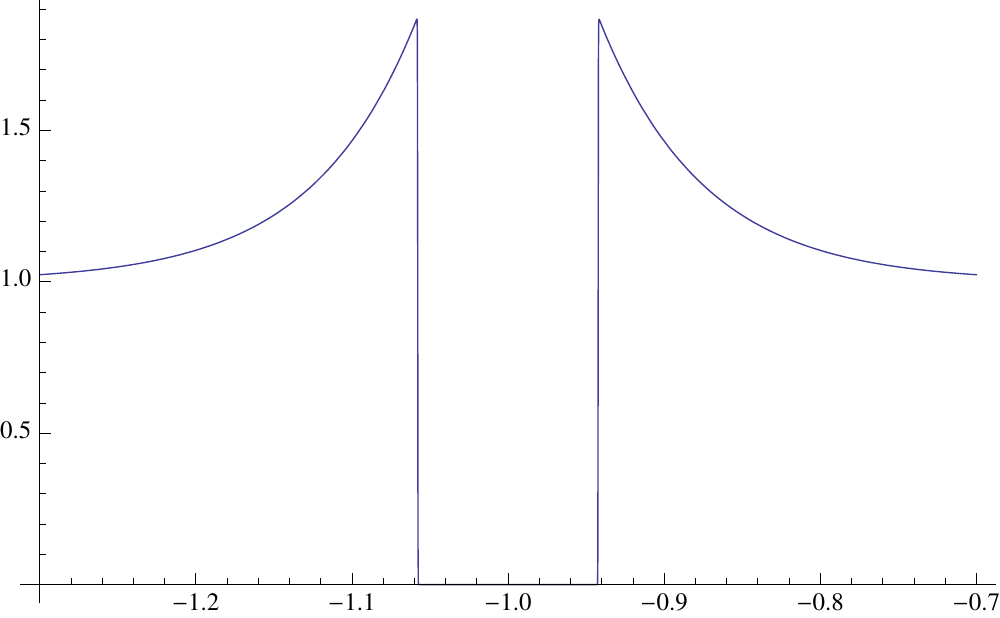} & \includegraphics[width=0.45\textwidth]{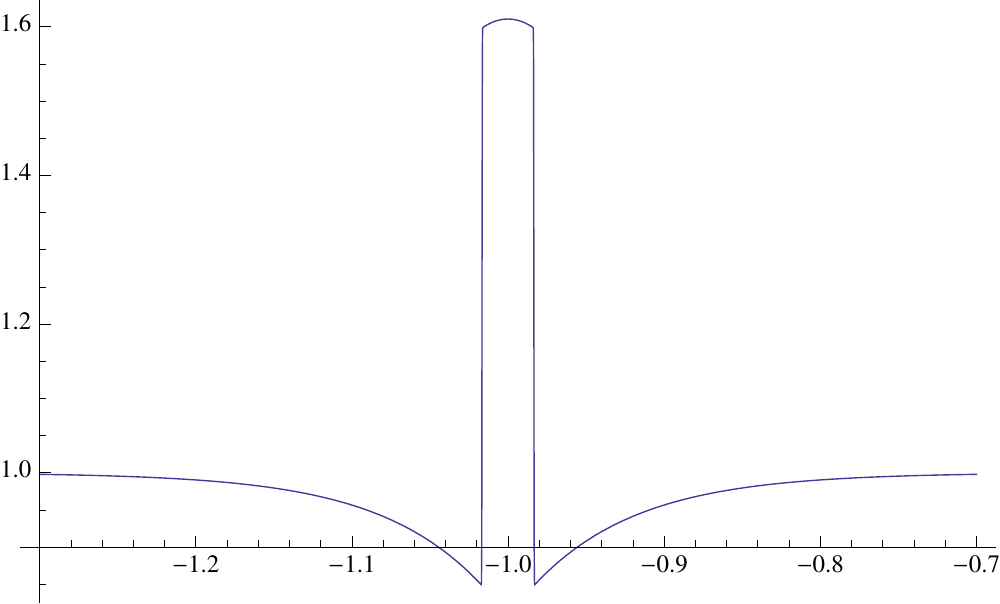} \\
	\hline
	\includegraphics[width=0.45\textwidth]{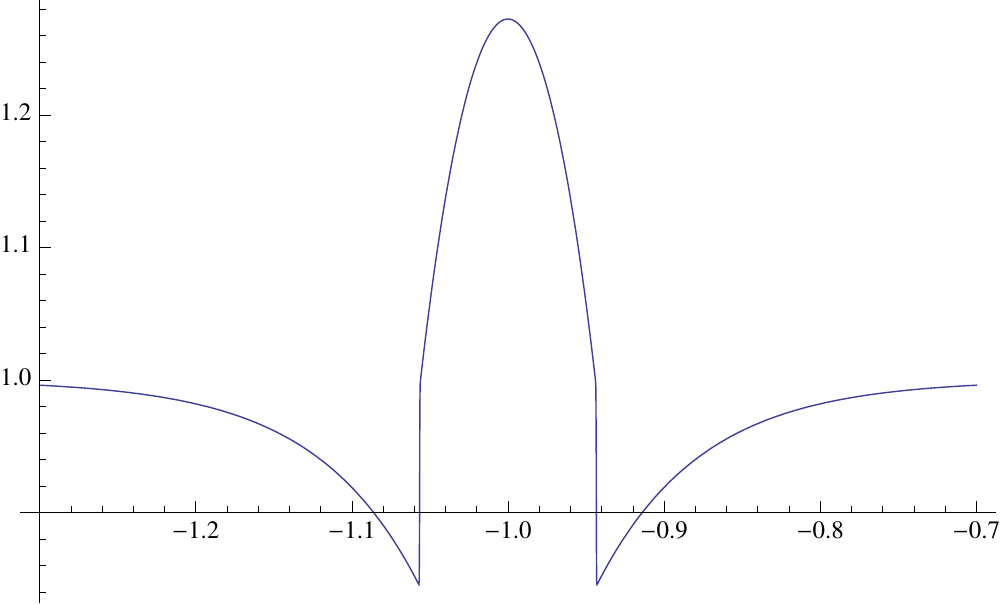} & \includegraphics[width=0.45\textwidth]{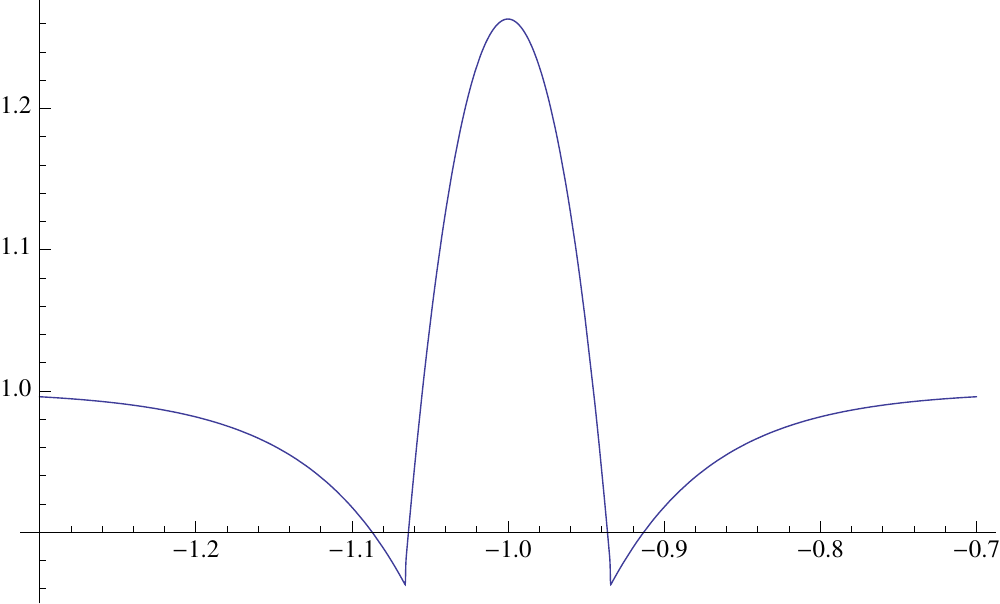} \\
	\hline
\end{array}$
\end{center}
\caption{\label{tablefp} The figures are snapshots of a symmetric collision at different times. In this set up, $c_1=c_2=1$, while the false vacuum is at $\phi=0$. The collision time is at $t_0=15$, and the figures show snapshots at times just before the collision ($t_1=8$), just after ($t_2=20$), a while after ($t_3=100$) and a long time after the collision ($t_4=1000$). We notice that the perturbation freezes (due to de Sitter background) and that the jump in $\phi$ disappears at late time leaving only a kink, a jump in the first derivative. Notice that the four figures are not at the same scale. }
\end{figure}

In \figref{tablefp} we show an example of a collision between two identical bubbles in the free-passage approximation. The figures in table \ref{tablefp} gives snapshots at different times of the field in (\ref{phitot}).

\subsection{Predictions for the temperature of the spot}

At least in models involving scalar fields, even minimal knowledge of the field space and the free-passage approximation can lead to a prediction for the sign (hot or cold) of the spot in the CMB temperature map.  
In the case of a collision between two identical bubbles, \eqref{freepass} shows that the initial perturbation always pushes the field away from the tunneling point.  
For example, in a collision between two bubbles of the right-hand minimum in \figref{landscape}, the analysis above suggests that the field value in region 3 will be $\phi_3 \sim 2 c_1$.  

In a model where $\phi_3 \sim 2 c_1 \simgeq \phi_{obs}$ ($\phi_{obs}$ is the minimum in the observer's bubble) the consequences for cosmology in region 3 are likely to be large and are unpredictable without more detailed knowledge of the model---the field could swing back across the minimum $\phi_{obs}$ and onto the inflationary plateau, or it could immediately reheat in the minimum.  In any case it seems unlikely that the affected region (region 3) will undergo \emph{more} efolds of inflation than the unperturbed region (region 1).
However, if the plateau is long compared to the barrier width, the field in region 3 may advance relative to the field in region 1 by the equivalent of only a few efolds---enough to have a potentially observable effect, but not necessarily a very large one.  The result will be fewer efolds of inflation in the affected region, since the perturbation pushes the inflaton forward down the inflationary plateau.
In both cases the result is  \emph{fewer} efolds of inflation in the affected region. After these results were obtained and while the manuscript was  in preparation, \cite{Johnson:2011wt} and \cite{Hwang:2012pj} appeared that perform numerical analyses in specific models that appear to support this conclusion.

Fewer efolds of inflation produces an intrinsic cold spot, heuristically because such a region reheats earlier and has longer to cool before last scattering.  Due to the Sachs-Wolfe effect this produces a \emph{hot} spot on the CMB (and a corresponding azimuthal polarization pattern around the edge of the disk \cite{Czech:2010rg}).  This conclusion is not model-independent---for instance, non-linear effects from interactions could invalidate the free-passage approximation---but we expect it to hold in a fairly wide range of models.  

By contrast, in a collision with a bubble containing another type of vacuum, the jump in the inflation can have the opposite sign.  A realistic model of the string landscape would contain a multi-dimensional field space.  Assuming $\phi_{3}$ is in the basin of attraction of the slow-roll inflationary trajectory, the effects are (tentatively) more likely to delay inflation in the affected region for the same reasons explained above.  In principle, they could also lead to interesting features in the power spectrum of perturbations in the affected region of the CMB sky---such as non-Gaussianities or oscillations---which would be produced as the field oscillates around the ``valley floor'' inflationary trajectory (a somewhat related phenomenon was recently studied in \emph{e.g.} \cite{Chen:2009we}).

\subsection{Domain wall boundary conditions}

An alternative approach is the one followed in \cite{wwc2}, where the domain wall separating the two bubbles after the collision was treated as a boundary condition for the inflaton.  The simplest boundary conditions are Dirichlet---that the inflaton field takes a fixed value on the domain wall.  Given the trajectory of the domain wall (which can be determined by treating it as a thin wall separating two regions of de Sitter space \cite{Freivogel:2007fx, wwc}), this fully determines the evolution of the inflaton field everywhere inside the observation bubble.  The initial conditions for the inflaton prior to the collision are set by the Coleman-de Luccia initial conditions along the lightcone of the nucleation point of the observer's bubble,  perturbed by the presence of the domain wall (\figref{delta}).  Together, the Coleman-de Luccia initial conditions plus the domain wall form a complete Cauchy surface for the evolution of the inflaton.

The inflaton perturbation during slow-roll inflation is determined by the functions $f(\xsi_{-})$ and $g(\xsi_{+})$ of \eqref{pert}.  Consider solving the boundary value problem described in the paragraph above.  The inflaton perturbation $\delta \phi$ due to the collision is zero outside the collision lightsheet.  Because the coordinate $\xsi_{-}$ is constant along null rays crossing the collision lightsheet, this means the function $f$ must be zero or constant.  Therefore, the inflaton perturbation is determined by the single function $g(\xsi_{+})$. 

Another way to understand this argument is to note that the value of $\delta \phi$ at any point in the $\xsi_{+}, \xsi_{-}$ plane is determined by its values at any two points, one along each of the past directed null geodesics originating from the point.  Since the left-going past null geodesic terminates on the Coleman-de Luccia initial condition surface, it is only the right-directed past null geodesic that determines $\delta \phi$, and this in turn determines the function $g(\xsi_{+})$.

The above arguments demonstrate that in this scenario, the perturbation depends only on the coordinate $\xsi_{+}$ perpendicular to the lightsheet of the collision.  Since the perturbation is zero outside the lightsheet, in the thin wall approximation the function $g(\xsi_{+})$ must be proportional to $\theta(\xsi_{+}-\xsi_{0})$, as in \eqref{pert}.  The question of whether or not $\delta \phi$ is actually discontinuous---rather than simply having discontinuous first or higher derivatives---comes down to precisely what boundary condition one imposes at the domain wall.  While we will not present further details here, in various examples it is clear that for generic choices, $\delta \phi$ is itself discontinuous.  Put another way, because the initial conditions  form a Cauchy surface, requiring $\delta \phi$ to be continuous on the lightsheet overconstrains the problem.  Therefore, generically there is a non-zero discontinuity in $\delta \phi$ (which then inflates into a discontinuity in the first derivative as explained in detail in \secref{powerseries} and App.~\ref{seriesappend}).

Our analysis on this point differs from that of \cite{wwc2}, which required continuity of the perturbation along the lightsheet.  However, our conclusion---that the inflaton perturbation due to a thin-wall collision near the end of inflation has a discontinuity in the first derivative---is identical.  The sign of the effect in the case of Dirichlet boundary conditions depends on the value of the inflaton on the domain wall in a way that was discussed in \cite{wwc2}.

\section{Thickness of the wall}\label{thickwall}

In this section we will explore the consequences of relaxing the thin-wall approximation.  Specifically, we will study a scenario where the collision bubble has walls of finite thickness.  In such a case the perturbation $\delta \phi$ is no longer proportional to theta function, but instead to a smooth function related to the field profile of the collision bubble's wall when it formed in the parent false vacuum.

\begin{figure}
	\begin{center}
	\includegraphics[angle=90, width=\textwidth]{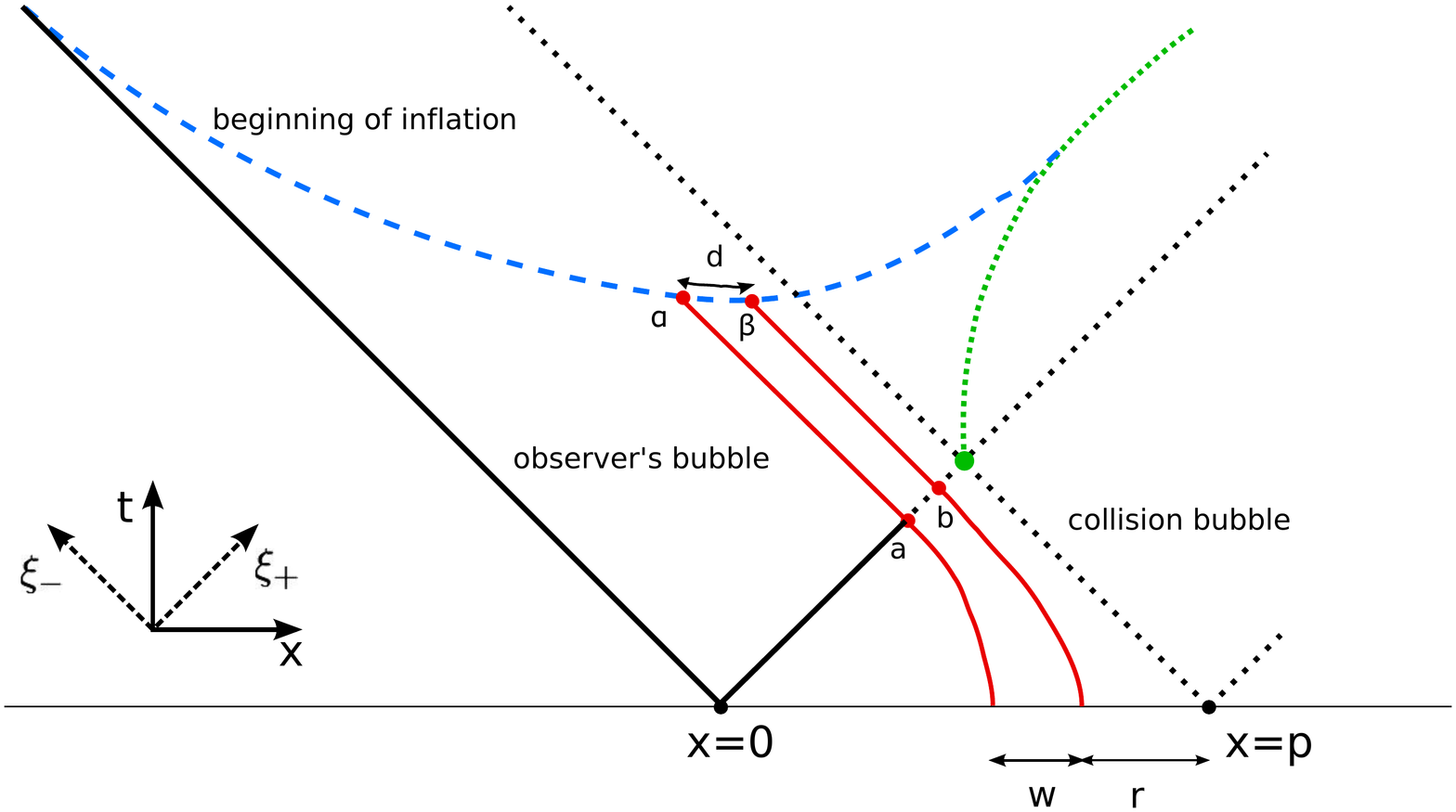}
	\caption{Colliding with a thick-wall bubble centered at $x=p$.  The red lines indicate the inner and outer edges of the thick wall and the region it affects.  As usual, each point in this diagram represents a 2-hyperboloid with a radius that depends on position in the plane of the diagram.}
	\label{wallthickness}
	\end{center}
\end{figure}

In our analysis we treat the thick wall as if it were composed of two thin walls separated by a distance $w$ when the collision bubble forms in the parent false vacuum, and determine the length scale $d$ corresponding to $w$ in the inflaton perturbation at the beginning of inflation (\figref{wallthickness}).  

We consider two distinct limits. Choosing a reference frame in which the two bubbles nucleate simultaneously and the observer's bubble is centered at $x=0$,  we first study the case where the collision bubble nucleates close to ours (at a distance $H_f p\ll 1$); then we consider the scenario in which the colliding bubble nucleates close to the de Sitter horizon of the false vacuum ($\pi - H_f p\ll 1$). 

In order for the collision to be visible, the lightsheet of the collision should intersect the last scattering surface of the observer.  That means that the lightsheet must intersect the FRW timeslice at the beginning of inflation near the comoving location of the observer.  As can be seen from Figures \ref{delta} and \ref{wallthickness}, when the collision occurs early as it does in our first limit ($p$ small), the lightsheet intersects the the beginning of inflation surface far to the left of $x=0$, and therefore an observer that can see it is boosted away from the collision bubble.  By contrast, if the collision bubble nucleates sufficiently close to the horizon of the parent false vacuum, the collision lightsheet may intersect the beginning of inflation surface near $x=0$ and hence could be visible to an observer at rest in this frame.

\subsection{Notation and setup}

We work with conventions illustrated in Fig. \ref{wallthickness}. Due to the symmetry of the Coleman-de Luccia initial conditions, prior to the collision all parts of the wall---defined for instance by contours of constant field value---will follow a hyperbolic (constant proper acceleration) trajectory in the parent false vacuum (curved red line segments in  \figref{wallthickness}).  Approximating the observer's bubble by the lightcone of its nucleation point for the moment, $a$ and $b$ mark the events where the inner and outer surfaces of the collision bubble's wall collide with ours.

After the collision, we propagate the effects along future directed null lightsheets (the straight red segments in \figref{wallthickness}).  These lightsheets intersect the beginning of inflation surface at the points marked $\alpha$ and $\beta$, separated by a distance $d$.  Our goal in this section is to determine $d$ from $w$, the collision bubble's radius $r$, and its center location $p$.  Because the spacetime curvature of our bubble is negligible until the beginning of inflation at a time of order $H_i^{-1} \gg H_f^{-1}$, we treat the interior of our bubble as Minkowski spacetime.

\subsection{Collision soon after nucleation}

 In this first limit, the two bubbles are nucleated at a distance $p\ll H_f^{-1}$ (Hubble distance of the false vacuum), the initial size of the bubble is $r_c$ and the thickness of the wall at nucleation is $w$.  In this limit, as explained above we can ignore the curvatures of both the parent false vacuum and slow roll inflation and treat the spacetime globally as flat Minkowski.

We set the coordinates in such a way that our bubble nucleates at $x=0$. Then the right branch of light cone of our bubble is given by $x=t$, and therefore the equations to solve to find $x_a$ and $x_b$ are
\begin{eqnarray}
	-x_a^2+(x_a-p)^2 & = & (r_c+w)^2\\
	-x_b^2+(x_b-p)^2 & = & r_c^2 ,
\end{eqnarray}
where we have accounted for the fact the the wall follows a hyperbolic trajectory. Even if at leading order we have $x_a\simeq x_b\simeq H_f^{-1}/2$, the difference is
\begin{equation}
	x_a-x_b\simeq\frac{r_cw}{p} \hspace{1mm} .
\end{equation}

We compute the coordinates of the point $\alpha$ ($\beta$ would be analogous) by finding the intersection between the hyperbolic surface at the end of inflation and the causal cone of $a$; namely we have to solve the system
\begin{eqnarray}
	-x_{\alpha}^2+t_{\alpha}^2 & = & H_i^{-2} \\
	x_{\alpha}-2x_a & = & -t_{\alpha} .
\end{eqnarray}

We find
\begin{eqnarray}
	x_{\alpha}&=&-\frac{H_i^{-2}-4x_a^2}{4x_a} \\
	t_{\alpha}&=&\frac{H_i^{-2}+4x_a^2}{4x_a}
\end{eqnarray}
and the result for $(x_{\beta},t_{\beta})$ would be the same with $a\leftrightarrow b$. Therefore
\begin{eqnarray}
	x_{\alpha}-x_{\beta} &=& (x_a-x_b)\left(1-\frac{H_i^{-2}}{4x_ax_b}\right) \\
	t_{\alpha}-t_{\beta} &=& (x_a-x_b)\left(1+\frac{H_i^{-2}}{4x_ax_b}\right).
\end{eqnarray}

The (hyperbolic) distance is now easily found:
\begin{eqnarray} 
	d^2 &=& (\Delta x)^2-(\Delta t)^2 \nonumber \\
	&=& H_i^{-2}\frac{(x_a-x_b)^2}{x_ax_b} , \label{hypdistM}
\end{eqnarray}
which can be written
\be
	d^2 \simeq H_i^{-2}\frac{(2r_cw+w^2)^2}{p^4-p^2(r_c^2+(r_c+w)^2)+r_c^2(r_c+w)^2} .
\ee

In order to compare with the computation we will show in the next section, it is useful to introduce the dimensionless parameter $\delta=H_fp$ (the distance between the bubble centers in units of $H_f^{-1}$):
\be
	d^2 = \frac{H_f^4}{H_i^2}\frac{(2r_cw+w^2)^2}{\delta^4-\delta^2H_f^2(r_c^2+(r_c+w)^2)+H_f^4r_c^2(r_c+w)^2}.
\ee
For our analysis to be valid the bubbles should not overlap when they nucleate, meaning that $p > r_c + w$.  Taking the limit $p \gg r_c,w$  gives
\be
	d H_i =  H_f^2 w(2r_c+w)\delta^{-2} ~~~~~~~~~~~~  (\delta \ll 1) . \label{distdelta}
\ee
The quantity $d H_i$ is the apparent wall thickness in units of the inflationary Hubble length; as such it roughly corresponds to the angular size of the apparent thickness on the CMB sky after minimal inflation.

\subsection{Collision with a bubble nucleated close to the de Sitter horizon}

When $p\ll H_f^{-1}$ the effects of the parent vacuum's curvature prior to the collision are negligible.  By contrast, when the bubble nucleates close to the horizon of the de Sitter  false vacuum ($\pi H_f^{-1}$) the effects of the de Sitter curvature become important. 

Again making use of the hyperbolic metric (\ref{hypmet}), the trajectories of the wall and the light cone of our bubble respectively are given by:
\begin{eqnarray}
	1+t^2 &=& \frac{\cos^2(r_c)}{\cos^2(x-(\pi-\epsilon))} = \frac{\cos^2(r_c)}{\cos^2(x+\epsilon)} \label{cwall}\\
	x&=&\tan^{-1}(t) \label{owall} ,
\end{eqnarray}
where for simplicity we set $H_f=1$ (we will restore $H_f$ restore at the end) and $\epsilon=\pi H_f^{-1}-p$ is the distance from the de Sitter horizon where the bubble nucleates. Also, from now on we might use $c(r_c)$ to indicate the constant $\cos^2(r_c)$.

Equating \eqref{cwall} and \eqref{owall} to find the collision point gives 
\be
	\frac{1}{\cos^2(x)}=\frac{c(r_c)}{\cos^2(x+\epsilon)}.
\ee
The point $x=\pi/2$ is midway between the center of the observer's bubble and the de Sitter horizon, and the divergence there indicates that a collision at that point would take place only after infinite time.  Since we are interested in collisions with bubbles that nucleate near the horizon, we expand in $\gamma=\pi/2-x \ll 1$ and $\epsilon \ll 1$. 
One finds
\be
	\frac{1}{\gamma^2} = \frac{c(r_c)}{\gamma^2+\epsilon^2-2\gamma\epsilon} .
\ee

This leads to only one physical solution once we impose the condition that $\gamma\to0$ when $r_c\to0$, so we find
\be
	\gamma = \frac{1-\cos(r_c)}{\sin^2(r_c)}\epsilon
\ee
and therefore
\beq
	x_b &=& \frac{\pi}{2} - \frac{1-\cos(r_c)}{\sin^2(r_c)}\epsilon \\
	t_b &=&\cot\left(-\frac{1-\cos(r_c)}{\sin^2(r_c)}\epsilon\right)
\eeq
and analogously one can find
\beq
	x_a &=& \frac{\pi}{2} - \frac{1-\cos(r_c+w)}{\sin^2(r_c+w)}\epsilon \\
	t_a &=&\cot\left(-\frac{1-\cos(r_c+w)}{\sin^2(r_c+w)}\epsilon\right) .
\eeq

We need now to match these results with the coordinates inside our bubble, that we treat as flat, and then determine the intersection points of the lightsheets with the surface at the beginning of inflation. Inside our bubble we use the hyperbolic slicing of Minkowski
\be
	ds'^2 = -dt'^2+dx'^2+t'^2dH_2'^2
\ee
so that the matching condition with the outside metric (\ref{hypmet}) is that the hyperboloids need to have the same radius on the light cone, where the two patches meet: $t'|_{lc}=t|_{lc}$. Therefore
\beq
	x_a' = t_a' &=& t_a \\
	x_b' = t_b' &=& t_b \nonumber .
\eeq

It is easy now to follow the null geodesics from these points and find their intersection with the beginning of inflation surface:
\beq
	-t'^2+x'^2 &=& H_i^{-2} \\
	x'^2 -2t_a &=& -t'
\eeq
giving
\beq
	x_{\alpha}' &=& \frac{H_i^{-2}}{4t_a}+t_a \\
	t_{\alpha}' &=& -\frac{H_i^{-2}}{4t_a}+t_a
\eeq
and
\beq
	x_{\beta}' &=& \frac{H_i^{-2}}{4t_b}+t_b \\
	t_{\beta}' &=& -\frac{H_i^{-2}}{4t_b}+t_b.
\eeq

Computing the hyperbolic distance on the surface at the beginning of inflation is straightforward:
\beq
	d^2 &= & H_i^{-2}\frac{(t_a-t_b)^2}{t_at_b} \nonumber \\
	 &=& \frac{H_f^4}{H_i^2}\frac{(2r_cw+w^2)^2}{64}\frac{(1+\cot(H_f\epsilon/2))^2}{\cot^2(H_f\epsilon/2)}H_f^2\epsilon^2 \nonumber \\
	  &\simeq& \frac{H_f^4}{H_i^2}\frac{(2r_cw+w^2)^2}{64}H_f^2\epsilon^2 \label{hypdist2}
\eeq
where we expanded for small $r_c$ and $w$ ($r_c,w\ll H_f^{-1}$) and in the last line we expanded for small $H_f^{-1}\epsilon$ as well.

Introducing again the parameter $\delta=H_fp$ the equation (\ref{hypdist2}) can be rewritten
\be
	d H_i \simeq H_f^2 w (2r_c+w )\frac{\pi-\delta}{8} ~~~~~~~~~~~~  (\pi-\delta \ll 1). \label{distpidelta}
\ee

Comparing with \eqref{distdelta} reveals the effects of the observer's boost.  When the bubbles nucleate very close together, the observer must be moving away from the collision bubble with a large Lorentz factor in order to be in position for the collision lightsheet to intersect her last scattering volume.  This boost away from the collision has the effect of Lorentz dilating the apparent thickness of the wall, which accounts for the negative power of $\delta$.  By contrast, if the collision bubble nucleates very close to the de Sitter horizon, the observer must be boosted towards it, and the thinner it appears.

Most bubble collisions will occur with a $\delta \simgeq 1$.  For such a case the apparent thickness is
\be
	d H_i \sim H_f^2 w (2r_c+w)/8 ~~~~~~~~~~~~  (\delta \sim 1).  \label{typdist}
\ee
For a thin wall bubble $H_f w \ll 1$, and in all cases $H_f (r_c+w) < 1$ (the critical size of the bubble must be less than one Hubble length).  

The results of \cite{Kleban:2011yc} showed that a thin-wall bubble produces a very interesting feature on the CMB sky at the degree scale (a ``double peak" in polarization intensity, for instance).  The quantity $d H_i$ is the ratio of the apparent width of the wall to the radius of curvature of the universe at the initial time.  Ratios of length scales are fixed during FRW expansion; therefore the right-hand side of \eqref{typdist} is also the ratio of the wall thickness to the curvature at the time of last scattering.  Observational constraints on curvature show that $|\Omega_k| \simleq .01$ in the current epoch \cite{Komatsu:2010fb}.  Using $\Omega_k = (a H)^{-2}$ and values from WMAP-7 \cite{Komatsu:2010fb}, the observational constraint combined with equation \eqref{typdist} shows that 
\be \label{rsh}
d_{LS}/r_{sh} \geq  300 H_f^2 w (2r_c+w)/8,
\ee
where $d_{LS}$ is the lengthscale corresponding to the wall thickness at last scattering, and $r_{sh}$ is the sound horizon.  The double peaks of \cite{Kleban:2011yc} are separated by an angular distance of order twice the sound horizon.  Therefore in order for the thin-wall approximation to accurately describe such a feature, the left-hand side of \eqref{rsh} should be less than one, and therefore
\be
(H_f w) H_f (2r_c+w) \simleq 10^{-2}.
\ee
 
 For collision bubbles with thicker walls, the thin wall approximation does not suffice to describe degree-scale features.  For instance, the double peak is likely to be strongly affected or smoothed over entirely.  Given the profile of the collision bubble's wall (which could be computed from the underlying microphysical model), a detailed analysis could be done to determine precisely what the effects on temperature and polarization are near the edge of the affected disk.  If there is substructure within the wall, it could have an interesting and non-trivial effect on cosmological observables in the region ``inside" the wall.  Conversely, the detection of a bubble collision in the CMB would allow one to infer something about the characteristics of the walls of the collision bubble, a remarkable fact considering its truly microscopic origin.

\section{Conclusion}\label{conclusions}

The detection of a cosmic bubble collision would be a discovery of tremendous importance.  As such, it is important to quantify its signatures accurately.  Because of the lack of a unique or easily calculable microphysical model, it is difficult to make entirely unambiguous predictions.  Nevertheless, we have shown here that the primary effects of bubble collisions in a broad class of models are parametrized by only a few parameters, and that this conclusion is robust under relaxing at least some of the assumptions made in previous analyses.  

One of the remarkable aspects of this problem is the potential for discovering certain details of near-Planck length physics by observing enormous structures that extend across the entire observable universe.  One example is the color (hot or cold) of the affected disk in the CMB temperature map, which we have argued is indicative of the type of vacuum in the collision bubble.  Another example is the thickness of the collision bubble's wall, which could imprint in an interesting way on the cosmological signature.  

One major direction that remains largely unexplored is to study a bubble collision in string theory, or in a model with at least some of its ingredients (compactified extra dimensions, fluxes, charged branes, etc.).  One obstacle is that a microphysical model for slow-roll inflation would be required, since the interactions of  the inflaton field are of major importance in determining the signatures.  We believe that the analysis of Section \ref{powerseries} is general enough to cover most such cases, at least when slow-roll inflation is effectively single-field.  However, in multi-field or more exotic inflation models interesting features might emerge.  Another potential signature is the effect of the collision on other light fields, which could potentially lead to perturbations in physical ``constants" such as the fine-structure constant $\alpha$ or probes of early universe physics such as 21cm \cite{Kleban:2007jd}.  Finally, recent work on ``barnacles" is interesting and should be further explored \cite{Czech:2011aa}.

\acknowledgments
We thank Guido D'Amico, Ben Freivogel, Thomas Levi, Marjorie Schillo, and Kris Sigurdson for useful discussions.
MK is supported by NSF CAREER grant PHY-0645435.  

\appendix
\section{Series expansions} \label{seriesappend}

In the analysis of \secref{powerseries}, we expanded the functions $f$ and $g$ in a power series in $x$ at an early time $\eta=\eta_0$ during inflation (Eqs. \ref{dphf0} and  \ref{dphg0}).
The coefficients in this expansion determine the initial perturbation $\delta \phi$ at $\eta=\eta_0$, and the full $\eta$ dependence of $f, g$ and $\delta \phi$ is trivial to determine from the coefficients $a_i$ and $b_i$ given \eqref{hypsol}.  However, under some circumstances it is more natural to instead expand the perturbation $\delta \phi$ at $\eta = \eta_0$ in a power series (for instance, if the initial perturbation is determined by the microphysics).  In this appendix, we determine the evolution of $\delta \phi$ in $\eta$ in terms of the coefficients in its expansion at $\eta_0$.

To begin, by analogy with a single term in \eqref{dphg0}, consider a perturbation that is a monomial in $x$ and zero outside the collision lightsheet:
\begin{equation} \label{eqexp}
	\delta \phi(x, \eta_0) =(x-x_0)^p\theta(x-x_0).
\end{equation}
Because only the left-moving part of the peturbation is observable, for simplicity we will set $f(\xi_-)=0$ (this condition implicitly determines the initial time derivative $\partial_\eta {\delta \phi}$).
Under these conditions, to find the evolution we wish to solve
\be
(x-x_0)^p\theta(x-x_0)=g_p(x+\eta_0)-\eta_0 g_p'(x+\eta_0)
\ee

The general solution, in terms of an integration constant, reads
\begin{eqnarray} \label{solfp}
	g_p(x+\eta_0) = e^{(x-x_0)/\eta_0} \eta_0^{p} \left(\Gamma(1+p)-\Gamma\(1+p;{C \over \eta_0}\)\right) \nonumber
	\\ -e^{(x-x_0)/\eta_0} \eta_0^{p} \left(\Gamma(1+p)-\Gamma\(1+p,{x-x_0 \over \eta_0}\)\right)\theta(x-x_0) .
\end{eqnarray}
Here $\Gamma(s;x) \equiv \int_x^\infty t^{s-1} e^{-t}dt$ is the incomplete gamma function.

There are two conditions \eqref{solfp} must satisfy. When we take the derivative, the term with the $\delta$ function arising from the differentiation of the Heaviside $\theta$ function must vanish: this is the case for all values of $C$. Second, the term which does not multiply the Heaviside $\theta$ function should vanish identically, because this solution should be non-zero only in the region affected by the collision (for $x+\eta>x_0+\eta_0$); this can be accomplished by choosing the integration constant $C=0$.

With $C=0$, the solution for all $\eta$ is
\begin{equation} \label{fexpsol}
	g_p(\xi_+) = (-)^{p+1}p!\left(e^{(\xi_+ - x_0 - \eta_0)/\eta_0}-\sum_{m=0}^p\frac{1}{m!}\left(\frac{\xi_+-x_0-\eta_0}{\eta_0}\right)^m\right) \theta(\xi_+ - x_0 - \eta_0),
\end{equation}
where as before $\xi_+=x+\eta$.
For notational simplicity, we can (by a shift in the coordinate $x$) choose $x_0=-\eta_0$, and will do so for the remainder of this appendix.

Consider the lowest order term $p=0$, which is the solution for the initial perturbation $\phi(x-x_0)=c_0\theta(x - x_0)$ for some constant $c_0$. The solution for $g$ is simply
\begin{equation}
	g_0(\xi_+) = c_0(1-e^{\xi_+/\eta_0}) \theta(\xi_+),
\end{equation}
and the field perturbation is
\begin{equation}
	\dph_{(0)}(\xi_+) = c_0(1-e^{\xi_+/\eta_0}+\frac{\eta}{\eta_0}e^{\xi_+/\eta_0})  \theta(\xi_+).
\end{equation}
For late times ($\eta \to 0$) this becomes
\begin{equation}
	\dph_{(0)}(\xi_+) \approx c_0(1-e^{\xi_+/\eta_0})  \theta(\xi_+).
\end{equation}

As we found in section \ref{powerseries}, an initial discontinuity in $\delta \phi$ inflates into a continuous function with discontinuous first derivative at the collision lightsheet.

In general, the initial condition at $\eta_0$ can be expanded
\begin{equation}
	\dph(x, \eta_0) = \sum_{n=0}^{\infty}c_n(x-x_0)^n \theta(x-x_0) .
\end{equation}
By linearity, the solution at general $\eta$ is then
\begin{eqnarray}
	\dph(x, \eta) &=& \sum_{n=0}^{\infty}c_n(-)^{n+1}n!\left[e^{\xi_+/\eta_0}-\sum_{m=0}^n\frac{1}{m!}\left(\frac{\xi_+}{\eta_0}\right)^m \right. \nonumber \\
	&& \left.-\frac{\eta}{\eta_0}\left(e^{\xi_+/\eta_0}-\sum_{m=0}^n\frac{1}{(m-1)!}\left(\frac{\xi_+}{\eta_0}\right)^{(m-1)}\right)\right] \theta(\xi_+) \nonumber \\
	&\approx& \sum_{n=0}^{\infty}c_n(-)^{n+1}n!\left[e^{\xi_+/\eta_0}-\sum_{m=0}^n\frac{1}{m!}\left(\frac{\xi_+}{\eta_0}\right)^m\right] \theta(\xi_+),
\end{eqnarray}
where the last line is valid as $\eta \to 0$.

\section{Spatial curvature}\label{curvature}

As mentioned in \secref{approxsec}, the curvature of the collision lightsheet becomes small in after inflation.  In fact, we will see that its radius of curvature is always larger than the radius of curvature of the FRW slice.

Using the coordinates (\ref{hypmet}) the radius of curvature of the lightsheet is simply $t$.  The radius of curvature of the FRW slice at the beginning of inflation is most easily found using a different set of coordinates:

\beq \label{dstrans}%
\cosh (H_i \tau) &=& \sqrt{1+(H_i t)^2} \cos( H_i x) ,\\
\sinh(H_i \tau) \sinh\beta \cos \theta &=& \sqrt{1+(H_i t)^2} \sin (H_i x) , \nonumber \\
\sinh(H_i \tau) \cosh \beta &=& H_i t \cosh\rho , \nonumber \\
\sinh (H_i \tau) \sinh \beta \sin \theta &=& H_i t \sinh \rho , \nonumber \\
\varphi &=& \varphi \nonumber.%
\eeq%
In these coordinates the metric (\ref{hypmet}) becomes
\beq \label{H3}
ds^2 &=& -d\tau^2 + H_{i}^{-2} \sinh^2 H_{i} \tau dH_3^2 , \\
dH_3^2 &=&  d\beta^2 + \sinh^2 \beta (d\theta^2 + \sin^2 \theta d\varphi^2) .
\eeq%
In these coordinates, after $N$ efolds inflation ends homogeneously at time $\tau_e \sim N H_i^{-1}$, and the radius of curvature of the universe at that time is $\sinh (H_i \tau_e)$.
We wish to compare this to the curvature of the lightsheet at the same time.  To do this, we should first find the intersection of the collision lightsheet---which is well approximated by the lightcone of a bubble that nucleated at $x=x_0$---with the hypersurface $\tau=\tau_e$. Using \eqref{dstrans}:
\beq
	& \sqrt{1+(H_i t)^2} & \cos(H_i x)  =  \cosh(H_i \tau_e)\simeq\frac{e^{H_i \tau_e}}{2} \label{infl1} \\
	& \sqrt{1+(H_i t)^2} & \cos (H_i (x-x_0)) =   1 \label{lc1}
\eeq
where we used the approximation $\tau_e H_i = N \gg 1$.

Taking the ratio of the two equations above one gets
\be
	\frac{\cos (H_i x)}{\cos(H_i (x-x_0))}\simeq\frac{e^{H_i \tau_e}}{2} .
\ee
Since $t \geq \tau \gg H_i^{-1}$ for all $x$ (\emph{c.f.} \eqref{dstrans}),  equation (\ref{lc1}) gives $t\simeq H_i^{-1}/\cos (H_i (x-x_0))$ and therefore
\be \label{tcurv}
	t\simeq\frac{H_i^{-1}e^{H_i \tau_e}}{2\cos (H_i x)} \simgeq H_i^{-1}e^{H_i \tau_e},
\ee
showing that the radius of curvature of the collision lightsheet at the end of inflation is of order or larger than the radius of the universe at that time.  The  denominator $\cos (H_i x)$ goes to zero when $x\sim\pm H_i^{-1} \pi/2$, which happens when the collision bubble nucleated very close to the horizon of the false vacuum $(+)$ or very close to the observer's bubble at $x=0$ $(-)$.

\bibliographystyle{klebphys2}

\bibliography{bubbles}

\end{document}